# Holistic Approach to the Periodic System of Elements

N.N.Trunov[*]

*D.I.Mendeleyev Institute for Metrology*

*Russia, St.Peterburg. 190005 Moskovsky pr. 19*

(Dated: February 20, 2009)

**Abstract:** For studying the objectivity and the quality of a given form of the Periodic system as a single whole we compare plots of functions presenting properties of elements in pairs of periods. Using mathematical statistics we introduce a dimensionless parameter which indicates high quality of the long form of the Periodic system.

## 1. Introduction

Since the discovery of the Periodic System of Elements (PSE) by Mendeleyev in 1869, it has been the object of extensive research by scientists of many specialties in its fundamental and applied aspects. Chemists treat the PSE as the basic for making new substances with preset properties. More than 400 geometric forms of the PSE are proposed and many suppositions are made but only at the level of qualitative observations and trends, see review [1]. On the other hand, a lot of partial conclusions and suppositions were made in order to connect forms of the PSE with the electron structure of the atoms by means of some qualitative use of the quantum mechanics. Unfortunately a great part of them is unsatisfied, see section 4.

Meanwhile, there is an obvious interest in solving the inverse problem: beginning with the information contained in the modern PSE to find numerical parameters characterizing the PSE as a single whole. Such holistic approach was announced and realized at the microscopic level in [2].

---

[*] Electronic address: trunov@vniim.ru

In the present paper we mainly concentrate attention on the macroscopic level, taking into account known physical and chemical properties of the elements. Note that holistic parameters founded and discussed below may be treated as metrological characteristics of the PSE, so that we unite and continue two main directions of Mendeleyev's activity: on the one hand – chemistry in general and the PSE specifically, and metrology in itself on the other hand.

## 2. Parameters evaluating the closeness of properties

As it follows from its name itself, the main feature of the PSE is its periodicity. We interpret it in such manner that plots of functions presenting each property must be similar in different periods. Hereafter we restrict ourselves to the neighboring pairs of periods since we suppose they have maximum similarity.

Now we have a) to introduce numerical parameters evaluating this similarity; b) to calculate these parameters for a given subset of properties and c) to exclude the possibility of random, but erroneous evidence of the closeness of properties.

We denote numerical values of properties as $p_n(Z)$, where $n$ is property number and $Z$ is the ordinal number of element. We can get rid of property-functions by including in the set $p_n$ coefficients for the change in properties with a change of external conditions (pressure, temperature and so on). The chosen subset of properties we denote as $\Pi = \{p_n\}$.

Different properties correspond to different measurement scales [3], i.e. have different permissible transformations of units measure. All the introduced parameters must be invariant under such transformations.

Most popular are scales of intervals and of relations. The first one permits linear transformations

$$p \to p' = ap + b \tag{1}$$

with any $a > 0$ and $b$ (e.g. measuring of time in different units and with a shift of beginning). The second one only permits (1) with $b = 0$ like the absolute temperature.

We require that $a$ may change independently in each of two compared periods. Let's denote $Z(N,M)$ an element of the long form of the PSE which is placed in the $N$–th period and in the $M$-th column. Hereafter we shall compare $p_n(Z)$ for each $n$ and pair of elements $Z(N,M)$ and $Z(N',M)$ in different periods but in the same column. Note that different periods may have different number of elements $L(N)$, so that for such periods we only can compare $\min(L(N), L(N'))$ pairs and some places in the PSE remain empty.

Hereafter we use the following parameter, which is invariant under transformations (1):

$$R_n(N,N') = R_n(N',N) = \sum_M t_n(N,M)\, t_n(N',M) \; ; \qquad (2)$$

$$t_n(N,M) = \frac{p_n[Z(N,M)] - \overline{p_n}(N)}{D_n(N)} \; , \qquad (3)$$

$$D_n^2(N) = \sum_M \left[ p_n(N,M) - \overline{p_n}(N) \right]^2 \; , \qquad (4)$$

where $\overline{p_n}(N)$ is the average value of $p_n$ over all $M$ in (2) or (4). Since $t_n(N,M)$ are normalized vectors,

$$-1 \leq R \leq 1 \; , \qquad (5)$$

an arbitrary function of $R$ also remains invariant under the transformation (1), but namely (2) is the most convenient one since it coincides with the well known coefficient of correlation and many statistical methods for its evaluation are known, see below.

Note that besides two discussed scales (1) there also exists the absolute scale with no transformation and all the sequential scales which permit any transformations maintaining ordering. For both these scales $p_n$ are dimensionless. A part of elements may not have a property $p_n$, so that in this case we get a partly ordering set. Hereafter we do not deal with these scales (though $R$ and $F$ (6) themselves belong to the absolute scale).

### 3. Parameters evaluating the quality of the PSE as a whole

Values of $R$ (2) - (4) for pairs of the neighbouring periods and for the indicated subset $\Pi$ of properties [4,5] are presented in the Table:

Table: Correlation coefficients $R$ (per cent) for the neighbouring pairs of periods

| Properties | 2-3 | 3-4 | 4-5 | 5-6 |
|---|---|---|---|---|
| Ionization energy | 93 | 96 | 85 | 97 |
| Energy of atom formation | 93 | 98 | 85 | 97 |
| Enthalpy of formation in the gas phase | 99 | 99 | 98 | 85 |
| Enthalpy of evaporation | 99 | 97 | 85 | 97 |
| Density | 77 | 82 | 97 | 98 |
| Boiling temperature | 94 | 98 | 85 | 98 |
| Melting temperature | 97 | 64 | 92 | 98 |
| | Min $R \geq 0.82$ | $L=8$ $\alpha \leq 0.01$ | | $L=18$ $\alpha < 10^{-4}$ |

Now we have to exclude the possibility of random, but erroneous evidence of the closeness of properties in different periods. Several methods of the mathematical statistics [6] give close significance levels $\alpha$ of the hypothesis for the absence or presence of correlation at given $R$ and min $L$. As it follows from the Table, the probability $\alpha$ of the absence of correlation is negligibly small for almost all calculated $R$.

If we need to have a sole parameter evaluating the objectivity of a given structure of the PSE with respect to a given subset $\Pi$ of properties it is meaningful to introduce the fraction $F(\Pi;\alpha)$ of all values $R$ in the Table for which the correlation does exist with the probability not less than $1-\alpha$. Evidently $0 \leq F \leq 1$. Thus for the Table :

$$F(\Pi;0.01) = 26/28 = 0.93 \ . \tag{6}$$

It is interesting to compare $F$ (6) with the same parameter for the same properties, but calculated for all pairs of periods $2 \leq N \leq 6$. Using results of [5], we get:

$$F(\Pi;0.01) = 55/70 = 0.79 \ .$$

Some change for the worse is mainly connected with irregularities of melting temperature.

In a similar way we can certainly compare other subsets of elements, e.g. the neighbouring groups at a fixed number of periods.

## 4. Discussion

The following fact attracts our attention in the Table: correlation coefficients $R$ are especially close to unity and respectively $\alpha$ is very small for the microscopic properties – ionization energy and energy of atom formation. It makes us to suppose a close connection between the structure of the PSE and the atomic structure. This supposition is really correct, namely a new effective quantum number [7]

$$T = \left(n + \tfrac{1}{2}\right) + \varphi\left(l + \tfrac{1}{2}\right) \tag{7}$$

may be introduced, where $n$ and $l$ are the radial and orbital quantum numbers respectively. New shells open and could be filled in order of increasing $T$ at the energy $\varepsilon = 0$. The new parameter $\varphi$ does not change with increasing amplitude of the self consisted potential with increasing $Z$. In such a way the genuine structure of the electron shells in atoms and thus the genuine structure of the PSE arises when the value of $\varphi$ is in the following interval [7]:

$$5/3 < \varphi < 2,$$

but a number of arguments allows to estimate $\varphi$ as:

$$\varphi = 1.72 \pm 0.05 \ .$$

Since $\varphi$ determines both the electron structure of each atom and of the PSE as a single whole we can imagine an abstract object – a "superatom" whose structure is governed by the only constant $\varphi$. The states of this "superatom" distinguished by Z are the ground states of an atom with Z electrons. ( Note that usual "explanations" of the PSE remain still valid outside this interval, e.g. for φ=1.6 , while shells ordering and the whole structure of the PSE changes drastically. That is why these qualitative arguments mentioned in handbooks are unsatisfied.)

Thus we have a pair of quite different holistic parameters $(F,\varphi)$ with F (6) determining the PSE as a single whole from the macroscopic and microscopic points of view respectively. It is however extremely difficult to trace how a given shells ordering ($\varphi$) results in a very high similarity of macroscopic properties (F) by means of atomic physics, electrodynamics, statistical physics and so on.

For the completeness of our approach it'll be interesting to calculate R and Π for each deformed structure of the PSE and then prove that the maximum value of Π is reached for the genuine long form of the PSE. One may expect that this really takes place since any deformation of the PSE disturbs extremely good correlation of the microscopic properties connected with the electron shells.

It should be mentioned in conclusion that similar holistic (or system) approach is also unavoidable for many new objects of great interest: clusters and other nano-objects and nano-systems, logic of the future quantum computers and the swarm calculations and so on. Many algebraic objects (graphs, matrices etc) may be used to describe ordering and other details of such systems.

As to the future development of metrology, though precise and reliable measurements still remain actual, its mainstream in the XXI-st century will be connected namely with holistic or system approach and with new types of metrological quantities.